\documentclass[preprint, amssymb,  aps, showpacs,preprintnumbers,
amsmath,showkeys,floatfix]{revtex4}

\usepackage{epstopdf}
\usepackage{graphics}
\usepackage{graphicx}
\usepackage{dcolumn}
\usepackage{bm}
\usepackage{longtable}
\usepackage{epsfig}
\usepackage{times}
\usepackage{url}
\usepackage{color}

\begin{document}

\title{Coherence enhanced optical determination of the $^{229}$Th isomeric transition}

\author{Wen-Te \surname{Liao}}
\email{Wen-Te.Liao@mpi-hd.mpg.de}

\author{Sumanta \surname{Das}}
\email{Sumanta.Das@mpi-hd.mpg.de}

\author{Christoph~H. \surname{Keitel}}
\email{Keitel@mpi-hd.mpg.de}

\author{Adriana \surname{P\'alffy}}
\email{Palffy@mpi-hd.mpg.de}

\affiliation{Max-Planck-Institut f\"ur Kernphysik, Saupfercheckweg 1, D-69117 Heidelberg, Germany}
\date{\today}


\begin{abstract}
The impact of coherent light propagation on the excitation and fluorescence of thorium nuclei in a
crystal lattice environment is investigated theoretically. We find that in the forward direction  
the fluorescence signal exhibits characteristic intensity modulations dominated by an orders of magnitude faster, sped-up initial decay signal.  This feature can be
 exploited for the optical determination of the isomeric transition energy. In order to obtain a unmistakable signature of the 
isomeric nuclear fluorescence, we put forward a novel scheme for the direct measurement of the transition 
energy via electromagnetically modified nuclear forward scattering involving two fields that couple three nuclear states. 
\end{abstract}

\pacs{
06.30.Ft, 
23.20.Lv, 
42.50.Gy, 
82.80.Ej  
} 
\maketitle

While atomic clocks based on the microwave ground state hyperfine transition 
of $^{133}$Cs \cite{Confi} have revolutionized the field of time-frequency metrology, 
general agreement exists that the next two orders of magnitude in precision pose 
serious challenges \cite{Cam12}. To circumvent this problem, an exciting alternative is to develop clocks based on a nuclear transition. The narrow line-widths of nuclear transitions and the isolation from external perturbations  promise
amazing stability. The optical transition from the ground state to the  isomeric, i.e., long lived, first excited 
state of $^{229}$Th was originally proposed by Peik and Tamm in 2003 \cite{Pei03} as a suitable candidate for a nuclear clock. Out of the entire known  nuclear chart, the $^{229}$Th ground state doublet consisting of the ground state $^{229g}$Th and the isomeric state  $^{229m}$Th offers a unique transition well within vacuum ultraviolet (VUV)  with energy below
10 eV \cite{Bec}. This transition is spotlighted also due to its tempting potential for testing the temporal 
variation of the fine structure constant \cite{Fla06}, building a nuclear laser in the optical range  \cite{Tka11}, or providing an exciting platform for  nuclear quantum optics  and coherent control \cite{Olga99,NQO2006,Stirap11,xpm12} of VUV photons. A key obstacle on the way  is the uncertainty about the transition energy: measurements up to date have been only indirect, suggesting an energy of $\sim 7.8\pm 0.5$ eV \cite{Bec}.

One intensely pursued approach for a more precise measurement of the $^{229}$Th isomeric transition and the realization of a nuclear frequency standard involves  nuclear spectroscopy of a macroscopic number of thorium nuclei doped in VUV-transparent crystals \cite{Rel10, Kaz12}. Due to the high doping density of up to $10^{18}$ Th/cm$^3$, crystals such as  LiCaAlF$_{6}$ \cite{Rel10} or CaF$_{2}$ \cite{Kaz12} offer the possibility to increase the  nuclear excitation  probability,  an important requirement especially considering the very narrow  radiative width of the transition, at present estimated at approximately 0.1 mHz \cite{Tka11}. Both  LiCaAlF$_{6}$ and CaF$_{2}$ have large
band gaps and present good transparency at the probable transition wavelength, such that the interplay with 
electronic shells in processes such as the electronic bridge \cite{eb2010}, internal conversion \cite{newprl} or nuclear excitation via electron capture or transition \cite{NEECT} can be neglected. The disadvantage of the crystal lattice environment is that inhomogeneous broadening due to temperature-dependent shifts of
the transition energy and in particular spin-spin relaxation compromise the Rabi \cite{Riehle2005}, Ramsey \cite{Riehle2005,Boyd2006} or hyper-Ramsey \cite{Yudin2010} interrogation schemes, commonly used in high-performance frequency standards. Instead, clock interrogation by fluorescence spectroscopy was proposed \cite{Kaz12}. Significant suppression of the inhomogeneous broadening is expected as long as all nuclei experience the same crystal lattice environment and are confined to the Lamb-Dicke regime, i.e., the recoilless transitions regime \cite{Dick53,Rel10}. However, together with the broad-band nature of the excitation, these very conditions lead to coherent light propagation through the sample and enhanced transient fluorescence in the forward direction, with a speed up of the initial decay depending primarily on the sample optical thickness.  These effects are well known from nuclear forward scattering  of synchrotron radiation (SR) \cite{vanB94,Smir96} driving M\"ossbauer nuclear transitions in the x-ray regime, and from the interaction of atomic systems with visible and infrared light \cite{crisp,HL83,R84}, but have so far never been addressed in the context of the $^{229}$Th isomeric transition.

In this Letter we investigate the effect of  coherent light propagation on the excitation and fluorescence signal of the VUV isomeric transition in $^{229}$Th.
We find that the  line-width of the light scattered in the forward direction is dynamically broadened by several orders of magnitude ($\sim 10^{6}$) due to the coherent decay of the large number of $^{229}$Th nuclei in the crystal and the strong dispersion close to the resonance frequency. While the dynamical fluorescence in the forward direction 
would therefore not be suitable for clock interrogation, it offers an enhanced and faster signal, advantageous for both a high signal-to-noise ratio and a shorter detection time.
In order to exploit the full potential of the dynamical broadening we put forward a quantum optics scheme 
based on quantum interference induced by two coherent fields coupling  three nuclear levels as a novel way 
to identify the isomeric transition energy. The proposed setup reminding of electromagnetically induced transparency (EIT) provides a clear signature for the excitation of the nuclear transition and enhanced precision in the optical determination of the transition frequency compared to a direct fluorescence experiment using only one field.

The reduced transition probability of the nuclear magnetic dipole  ($M1$)  $^{229}$Th isomeric transition has been evaluated theoretically to $B(M1)\simeq 0.032$ Weisskopf units \cite{Tka98,Tka11}. This value corresponds to the very narrow radiative transition width of approximately 0.1 mHz. Since such narrow-bandwidth sources are not available in the VUV region, broadband excitation, with either synchrotron or VUV laser light, is envisaged to be employed in order to resonantly drive the isomeric transition. As long as elastic, recoil-free scattering of the incident light occurs, the contributions of all nuclei are spatially in phase in the forward direction and interfere coherently \cite{vanB99}. To the forward signal contribute not only the resonant frequency components, which are absorbed and reemitted, but also the non-resonant components, which experience dispersion. Thus the time evolution of the forward scattering response does not follow a natural exponential decay as expected for fluorescence involving a single-scattering event, but exhibits pronounced intensity modulations characteristic for the coherent resonant pulse propagation \cite{vanB99}. This modulation is known in nuclear condensed-matter physics under the name of dynamical beat and for a single-resonance and a short $\delta(t)$-like exciting pulse has the form \cite{crisp,kagan,bible} $E(t)\propto \xi \exp^{-\tau/2} J_1(\sqrt{4\xi\tau})/\sqrt{\xi\tau}$, where $E$ is the transmitted pulse envelope, $\tau$ a dimensionless time parameter $\tau=t/t_0$ with $t_0$ denoting the natural lifetime of the nuclear isomer, $\xi$ the optical thickness and $J_1$ the Bessel function of the first kind. The optical thickness is defined as $\xi=N\sigma L /4$ \cite{bible}, where $N$ is the number density of $^{229}$Th nuclei, $\sigma$ the resonance cross section and $L$ the sample thickness. For a dopant density of $10^{18}$ Th/cm$^3$ and a sample with $L=1$~cm, we obtain $\xi\simeq 10^6$.

From the asymptotic behavior of the Bessel function of first kind we see that for early times $\tau\ll 1/(1+\xi)$ immediately after the excitation pulse (chosen as $t=\tau=0$) the response field has the form $E(t)\propto \xi \exp^{-(1+\xi)\tau/2}$, showing the speed-up of the initial decay. A parallel has been established between this faster decay and the superradiant decay in single-photon superradiance \cite{Scu06}.  For the case of the  1 cm crystal doped with $^{229}$Th considered above, the speed-up would correspond to an enhancement of the decay rate by six orders of magnitude. This value significantly reduces the gap between the incoherent relaxation time of the excited state population (the natural radiative decay) and the short coherence time due to the crystal lattice effects, bringing inhomogeneous and dynamical broadening on the same order of magnitude.  At later times, the decay becomes subradiant, i.e., with a slower rate comparable to the incoherent natural decay rate  due to destructive interference between radiation emitted by nuclei located at different depths in the sample.  A symmetric excitation of the sample would lead to the creation of a timed Dicke state \cite{Ralf2010} and a prolonged superradiant decay. So far this has been achieved for $^{57}$Fe nuclei embedded in the center of a planar waveguide in the first measurement of the collective Lamb shift in nuclei \cite{Ralf2010}. One should however keep in mind that the dynamical beat behavior is by no means restricted to nuclear systems and the x-ray radiation wavelength, having been observed over a large wavelength range $\lambda\simeq 1$ \AA$-1$ m in a variety of systems (see \cite{vanB99} and references therein). 

\begin{figure}[b]
\vspace{-0.4cm}
  \includegraphics[width=10cm]{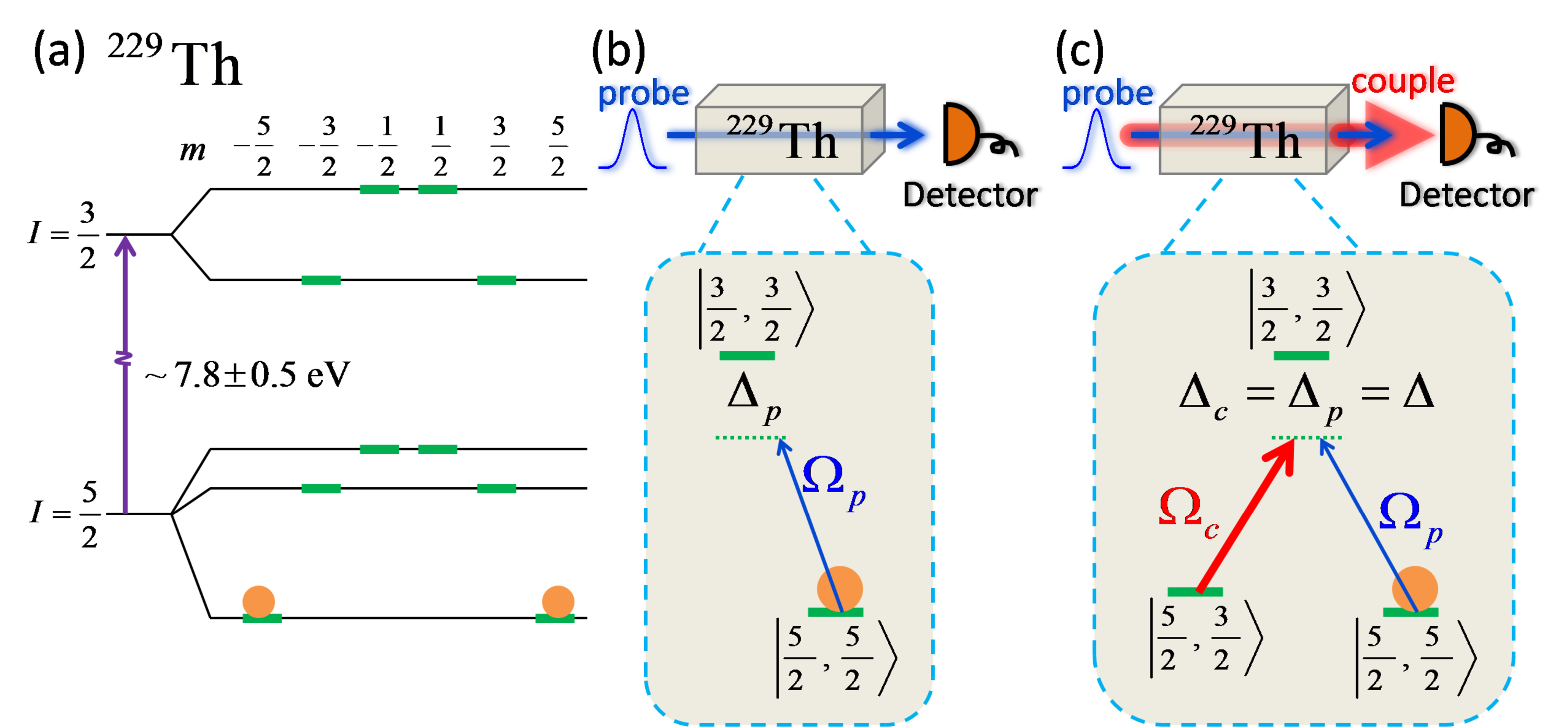}
  \caption{\label{fig1}
  (a) Quadrupole level splitting for the $^{229}$Th ground and isomeric states in the $^{229}$Th:CaF$_{2}$ crystal lattice environment. (b) A left circularly polarized probe field drives  the $|\frac{5}{2}\frac{5}{2}\rangle\rightarrow|\frac{3}{2}\frac{3}{2}\rangle$ transition.    (c) The electromagnetically modified forward scattering setup. The red thick arrow denotes the strong coupling field while the blue thin arrow shows the weak probe field. The detunings of the  probe and coupling fields to the respective resonance frequencies  are denoted by $\triangle_{p}$ and $\triangle_{c}$, respectively.
  }
\end{figure}

In the case of $^{229}$Th doped in VUV transparent crystals both conditions for (i) recoilless, coherent excitation and decay and (ii) broadband excitation are fulfilled for  scattering in the forward direction.  The incident laser or SR pulse duration is much shorter than the nuclear lifetime and provides broadband excitation. Due to the crystal transparency at the nuclear transition frequency, the main limiting factor for  coherent pulse propagation, namely electronic photoabsorption, is not present in this case. We are interested in the initial superradiant response which occurs on a time scale much faster than the incoherent radiative decay and can facilitate the optical determination of the isomeric transition energy. As a consequence of the hyperfine splitting affecting the $^{229}$Th nuclei in the crystal lattice, an analytical result for the scattered field intensity 
is not available and the numerical solution of the Maxwell-Bloch equations \cite{Scully2006} has to be evaluated instead.  This treatment allows also to take into account the inhomogeneous broadening occurring due to spin-spin relaxation in the crystal sample. We use in the following the quadrupole structure  with  hyperfine level 
energies given by $E_{m} \simeq Q_{is(g)}(1-\gamma_{\infty})\phi_{zz}[3m^{2}-I_{is(g)}(I_{is(g)}+1)]/
[4I_{is(g)}(2I_{is(g)}-1)]$ \cite{Abr61}, where $Q_{is(g)}$ = 1.8 $e$b (3.15 $e$b), with $e$ the electron charge,  is the quadrupole 
moment of the isomeric (ground) nuclear state, $\gamma_{\infty} = -(100-200) $ is the antishielding factor 
and $(1-\gamma_{\infty})\phi_{zz} \sim -10^{18}$ V/cm$^{2}$ electric field gradient \cite{Tka11, Bem88, Fei69}. 
Fig.~\ref{fig1}(a) shows the energy scheme of $^{229}$Th with the electric quadrupole splitting \cite{Tka11,Kaz12} of the ground and 
excited $^{229}$Th nuclear states of spins $I_g$=5/2 and $I_e$=3/2, respectively. Here we adopt the recently proposed 
$^{229}$Th:CaF$_{2}$ crystal \cite{Kaz12} under the thermal equilibrium condition \cite{Tka11}, i.e., all the nuclei are equally 
populating the two ground states of $m_{g}=\pm 5/2$ at sub-kelvin temperatures.

Plain fluorescence spectroscopy of the isomeric nuclear transition is plagued by a small signal-to-background 
ratio \cite{Rellergert2010b}. For both determination of the exact nuclear transition frequency and for the nuclear clock 
interrogation schemes, this is a major issue, especially due to the weak coupling of the nuclear transition to the radiation 
field. According to current experimental data \cite{Rellergert2010b}, the main source of background photons is the $\alpha$-decay 
of the $^{229}$Th ground state with the total counting rate on the order of  MHz and  $4\pi$ spatial distribution. A possible 
solution is offered by coherent light scattering, which ensures that a directional signal collecting in the forward direction 
benefits of a more favorable signal-to-background ratio. However, impurities and color centers (which unfortunately may occur 
during the experiment as irradiation effects) can cause residual absorption and may interfere with the $^{229}$Th optical 
spectroscopy even in the forward direction. This is why a clear signature of the nuclear isomeric transition 
$^{229g}$Th$\rightarrow$ $^{229m}$Th is desirable. In the following we put forward a quantum optics scheme designed to provide 
such a nuclear fluorescence signature by employing two VUV fields that drive a  three-level system in a setup reminding of 
EIT \cite{Scully2006}. Each field connects one of the two ground states to a third common excited state, a configuration commonly 
referred to as a three-level $\Lambda$ system \cite{Scully2006} in quantum optics.

A radiation pulse denoted in the following as  probe driving the relevant nuclear transition shines perpendicular to the nuclear sample, 
as shown in Fig.~\ref{fig1}(b).  We consider a left circularly polarized weak VUV probe field driving the $m_{e}-m_{g}=-1$ magnetic 
dipole transition, where  $m_{e}$ and $m_{g}$ denote the projections of the excited and ground state nuclear spins on the quantization 
axis, respectively. Additionally, as shown  in Fig.~\ref{fig1}(c), we apply a strong linearly polarized continuous wave (CW) denoted 
as coupling laser field driving the $m_{e}-m_{g}=0$ transition. A $\Lambda$-type scheme is formed due to the combination of initial 
nuclear population and the selected field polarizations. The purpose of applying the coupling field is to create the so-called dressed 
states \cite{Scully2006}, i.e., two linear combinations of the $|\frac{3}{2},\frac{3}{2}\rangle$ and $|\frac{5}{2},\frac{3}{2}\rangle$
nuclear states. This splits the nuclear resonance driven by the probe field into a doublet \cite{Shvydko1999H} via the 
Autler-Townes effect \cite{AutlerGlover}, giving rise to electromagnetically modified time spectra. The resulting beating  can be then 
used as a specific signature for the detection of the nuclear fluorescence photons.

We study the coherent pulse propagation for both setups in Fig.~\ref{fig1}(b) (assuming $\Omega_c=0$)  and Fig.~\ref{fig1}(c) by numerically determining the dynamics of the density matrix $\widehat{\rho}$ via the Maxwell-Bloch equations \cite{Shv99,Scully2006,Palffy2008,xpm12}:
\begin{eqnarray}
&&
\partial_{t}\widehat{\rho} = \frac{1}{i\hbar}\left[ \widehat{H},\widehat{\rho}\right]+\widehat{\rho}_{s}\, ,
\nonumber\\
&&
\frac{1}{c}\partial_{t}\Omega_{p}+\partial_{y}\Omega_{p}=i\eta a_{31}\rho_{31}\, ,
\label{eq1}
\end{eqnarray}
with the interaction Hamiltonian given by 
\begin{equation}
\widehat{H} = -\frac{\hbar}{2}\left(
\begin{array}{cccc}
  0 & 0 & a_{31}\Omega^{*}_{p}\\
  0 & -2(\Delta_{p}-\Delta_{c}) & a_{32}\Omega^{*}_{c}\\
  a_{31}\Omega_{p} & a_{32}\Omega_{c} & -2\Delta_{p}
\end{array}  
\right)\, ,
\nonumber
\end{equation}
and the decoherence matrix
\begin{equation}
\widehat{\rho}_{s} = -\left(
\begin{array}{cccc}
  -a^{2}_{31}\Gamma\rho_{33} & \gamma_{21}\rho_{12} & (\gamma_{31}+\frac{\Gamma}{2})\rho_{13}\\
  \gamma_{21}\rho_{21} & -a^{2}_{32}\Gamma\rho_{33} & (\gamma_{32}+\frac{\Gamma}{2})\rho_{23}\\
  (\gamma_{31}+\frac{\Gamma}{2})\rho_{31} & (\gamma_{32}+\frac{\Gamma}{2})\rho_{32} & \Gamma\rho_{33}
\end{array}  
\right).
\nonumber
\end{equation}
In the equations above  $\rho_{jk}=A_{j}A^{*}_{k}$ for $\{j,k\}\in \{1,2,3\}$ are the density matrix elements of $\widehat{\rho}$ for the nuclear wave function $|\psi\rangle=A_{1}|\frac{5}{2}\frac{5}{2}\rangle+A_{2}|\frac{5}{2}\frac{3}{2}\rangle+A_{3}|\frac{3}{2}\frac{3}{2}\rangle$ with the nuclear hyperfine levels shown in Fig.~\ref{fig1}(a). Furthermore,  $(a_{31},a_{32})=(\sqrt{2/3},-2/\sqrt{15})$ are the corresponding Clebsch-Gordan coefficients. The matrix $\widehat{\rho}_{s}$ describes the decoherence due to spin relaxation $(\gamma_{31},\gamma_{32},\gamma_{21})=2\pi\times(251,108,30)$ Hz \cite{Kaz12} and  $\Gamma=0.1$ mHz \cite{Tka11} denotes the spontaneous decay. The parameter $\eta$ is defined as $\eta=\frac{\Gamma\xi}{2L}$, and we consider a target thickness of  $L=1$ cm. Further notations are $\Omega_{p(c)}$ for the Rabi frequency of the probe (coupling) fields which is proportional to the electric field $\vec{E}$ of the probe (coupling) VUV radiation  \cite{Scully2006, Palffy2008} and $c$ the speed of light.

\begin{figure}[b]
\vspace{-0.4cm}
  \includegraphics[width=10cm]{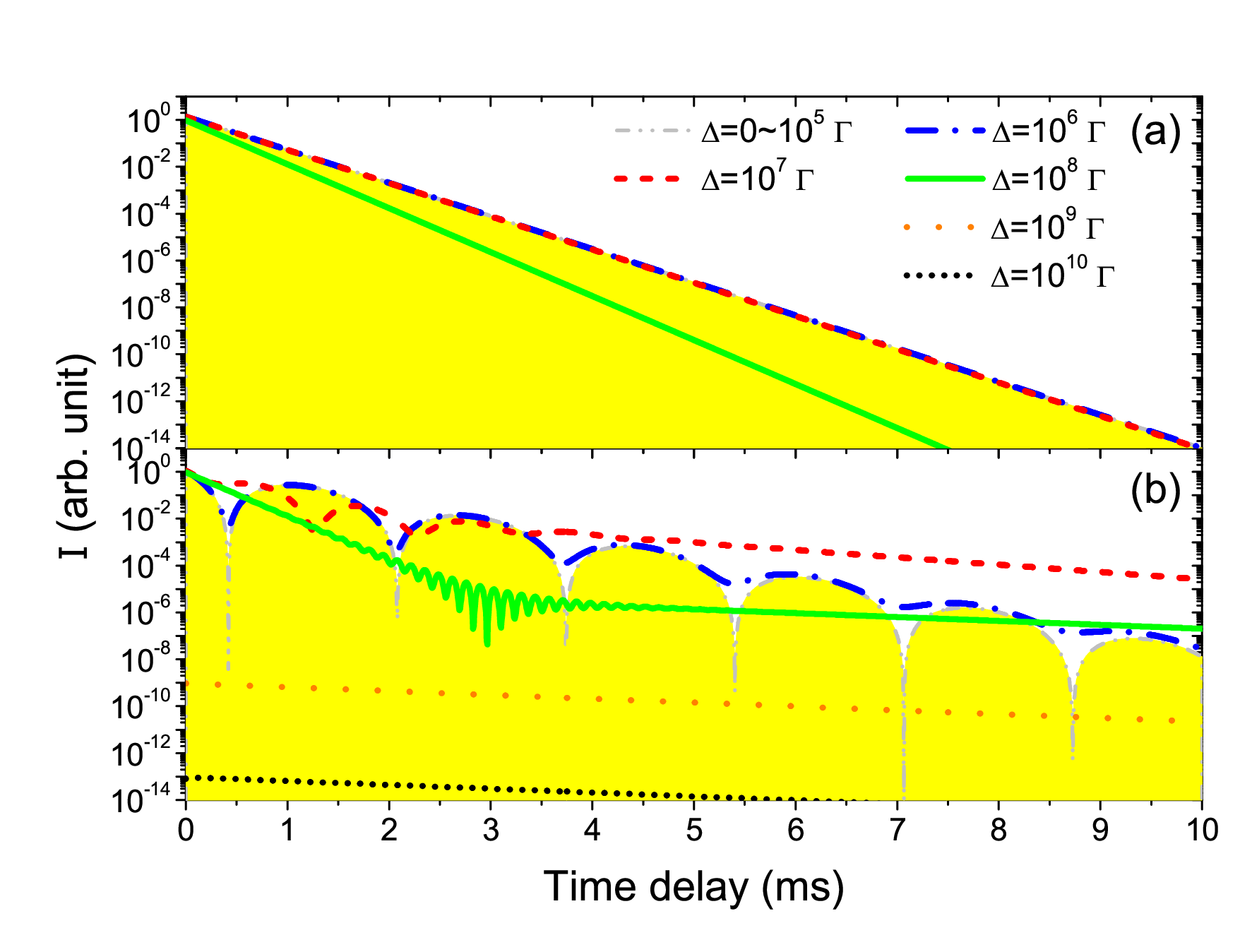}
  \caption{\label{fig2}
  The time spectra of the forward scattered signal (a) without coupling field, (b) with coupling field. For all spectra the detunings  $\triangle_{p}=\triangle_{c}=\triangle$   take values between $0\leq\triangle\leq 10^{10}\Gamma$. The yellow filled area below the gray dashed double-dotted line delimits the region  $0\leq\triangle\leq 10^{5}\Gamma$.  In (a) the values for  $\triangle=10^{9}\Gamma$ and $10^{10}\Gamma$ are smaller than the displayed scale.
  }
\end{figure}

Since the electric quadrupole splitting of ground states can be experimentally determined via standard nuclear magnetic resonance (NMR) techniques \cite{Kaz12}, we may set the two detunings of the coupling and probe fields to be identical but unknown initially, i.e., $\triangle_{c}=\triangle_{p}=\triangle$. We numerically solve Eq.~(\ref{eq1}) with $\xi=10^{6}$ and $\Omega_{c}$ with a laser intensity of 2kW/cm$^{2}$ and  scan the region $0\leq\triangle\leq 10^{10}\Gamma$. The time spectra  $\vert \Omega_{p}(t,L)\vert^2$ for the setup in Fig.~\ref{fig1}(b) are presented in Figs.~\ref{fig2}(a). In the absence of the coupling field, the probe only interacts with a two-level nuclear system, and the corresponding time spectra are less sensitive to the detuning $\triangle$. For $\triangle=10^{9}\Gamma$ and $10^{10}\Gamma$, the probe field will not create any nuclear excitation and just passes through the crystal. The $^{229}$Th nuclei will start to coherently scatter probe photons when $\triangle\leq 10^{8}\Gamma$, and with smaller detunings the time spectra  become identical for $\triangle\leq 10^{7}\Gamma$. Essentially we could approach the wanted energy of $^{229m}$Th with the precision of $10^{7}\Gamma$, i.e. about 1 kHz, by measuring the signal photons scattered in the forward direction. As a disadvantage, since the spectra do not present any specific feature to confirm the excitation of the isomer, background from other unwanted electronic processes may present an identical scattering pattern at probe laser frequencies far away from the nuclear resonance. 

We turn now to the two-field setup presented in Fig.~\ref{fig1}(c), which is more detuning-sensitive and provides specific identification features in the scattering time spectra. In Fig.~\ref{fig2}(b) we present our results for the electromagnetically modified forward scattering spectra. With the influence of $\Omega_{c}$ the spectra are much more sensitive to the detuning, and probe photons start to interact with the $^{229}$Th nuclei already at $\triangle\leq 10^{10}\Gamma$. The enhanced sensitivity of the electromagnetically modified setup in Fig.~\ref{fig1}(c) is due to the detuning-sensitive dispersion relation which leads to a unique time spectrum for each combination of input probe field and coupling field strength and detuning. A comparison between Figs.~\ref{fig2}(a) and \ref{fig2}(b) shows that the coupling field has introduced additional beatings in the spectrum which can serve as a clear signature of the nuclear excitation. Furthermore, a significant advantage of our scheme is that the shapes  of the spectra are not identical until $\triangle\leq 10^{5}\Gamma$.  A fit of the theoretical and experimental curves can therefore be employed to determine the nuclear transition frequency.
 By scanning the detunings $\triangle$, i.e., by varying the known frequencies of probe and coupling simultaneously,  several $\triangle$-dependent forward scattering time spectra can be measured. A fit with the theoretical curves in Fig.~\ref{fig2}(b) allows the determination of the detuning value and thus of the  $^{229g}$Th$\rightarrow$ $^{229m}$Th transition frequency down to a precision of $10^{5}\Gamma$, i.e., 10 Hz. Once the isomeric transition frequency has been identified, the fluorescence clock interrogation scheme using the incoherent response emitted at an angle to the excitation pulse direction can be employed for the nuclear frequency standard.

A brief estimate of the experimental counting rate for the signal and background photons in the forward direction is performed for the case of a $^{229}$Th:CaF$_{2}$ crystal with the size 3 mm$\times$ 3 mm $\times$ 10 mm and $^{229}$Th concentration of $10^{18}$ cm$^{-3}$. Given the 7880 yr half life of the ground state $^{229}$Th, and the assumed 0.3 background photon per $\alpha$-decay \cite{Rellergert2010b}, the total counting rate of background photons (in $4\pi$ solid angle) is then $0.75$ MHz, which could be significantly suppressed down to $1.8$ Hz by registering only signals within $1^{\circ}\times 1^{\circ}$ in the forward direction. The background generated by the VUV pulse itself in the forward direction can be eliminated with a chopper or employing the nuclear lighthouse effect \cite{RalfNLE}. Thus, $20$ Hz would suffice as  counting rate of the scattered probe photons in the forward direction, corresponding to a signal-background ratio  larger than 10. On the other hand, the spin relaxation makes the coherences decay according to $\rho_{31}\sim e^{-\gamma_{31}t}$ since $\gamma_{31}\gg\Gamma$ \cite{Kaz12}. The coherent effects should therefore be observed within the time scale of ms allowing $100\sim 500$ shots  per second.

A CW laser source in the VUV region is presently available only within limitations. The KBe$_2$BO$_3$F$_2$ (KBBF) crystals \cite{kbbf}  have been successful in generating narrow-band VUV radiation via harmonic generation owing to their wide transparency and large birefringence necessary for phase-matched frequency conversion processes in  this frequency region \cite{vuv_kbbf1,vuv_kbbf2}. In particular, a quasi-CW laser with a 10 MHz repetition rate and 20 ps pulse duration has been achieved \cite{vuv_kbbf2}. A CW coupling VUV laser at around 160 nm wavelength could also be generated by sum frequency mixing in metal vapors or driving a KBBF crystal with Ti:Sapphire laser \cite{nolting,togashi}.  The circularly polarized probe laser  on the other hand may be generated via nonlinear sum-frequency mixing \cite{irrgang}, or a harmonic of a VUV frequency comb \cite{jones,ozawa} around
the isomeric wavelength. In the absence of a CW coupling field for the electromagnetically modified forward scattering scheme, 
one can use a moderate external magnetic field ($B=1$ G) to split   two degenerate $\Delta m=0$ transitions, for instance 
$|\frac{5}{2},\pm \frac{1}{2}\rangle \rightarrow |\frac{3}{2},\pm \frac{1}{2}\rangle$, into a doublet due to 
the different hyperfine energy shifts introduced by the magnetic dipole interaction. Numerical results for the LiCAF crystal for which 
the calculation of the magnetic hyperfine splitting is straightforward show that the detuning sensitivity goes as far as $\Delta=10^7\Gamma$, 
i.e., one could determine the nuclear transition frequency within 1 kHz with this method. 

In conclusion, the  dynamical broadening of the nuclear signal in the forward direction can be used as efficient tool to determine the isomeric transition frequency. A two-field setup reminding of EIT provides both a signature of the nuclear excitation and enhanced sensitivity to the field detuning from the resonance. 
This scheme demonstrates a novel way to solve the question of identifying the isomeric transition energy and lays the foundation for developing nuclear quantum optics in $^{229}$Th.

The authors  gratefully acknowledge fruitful discussions with G. Kazakov and Y. Nomura.


\end{document}